\documentclass[12pt,epsf]{article}

\usepackage{epsfig}
\usepackage{amssymb}
\usepackage{graphicx}
\usepackage{color}

\begin{document}

\baselineskip 6mm
\renewcommand{\thefootnote}{\fnsymbol{footnote}}


\newcommand{\nc}{\newcommand}
\newcommand{\rnc}{\renewcommand}

\headheight=0truein
\headsep=0truein
\topmargin=0truein
\oddsidemargin=0truein
\evensidemargin=0truein
\textheight=9.5truein
\textwidth=6.5truein

\rnc{\baselinestretch}{1.24}    
\setlength{\jot}{6pt}       
\rnc{\arraystretch}{1.24}   



\newcommand{\tcb}{\textcolor{blue}}
\newcommand{\tcr}{\textcolor{red}}
\newcommand{\tcg}{\textcolor{green}}


\def\be{\begin{equation}}
\def\ee{\end{equation}}
\def\ba{\begin{array}}
\def\ea{\end{array}}
\def\bea{\begin{eqnarray}}
\def\eea{\end{eqnarray}}
\def\nn{\nonumber\\}


\def\ct{\cite}
\def\la{\label}
\def\eq#1{Eq. (\ref{#1})}


\def\a{\alpha}
\def\b{\beta}
\def\g{\gamma}
\def\G{\Gamma}
\def\d{\delta}
\def\D{\Delta}
\def\ep{\epsilon}
\def\e{\eta}
\def\ph{\phi}
\def\Ph{\Phi}
\def\ps{\psi}
\def\Ps{\Psi}
\def\k{\kappa}
\def\l{\lambda}
\def\L{\Lambda}
\def\m{\mu}
\def\n{\nu}
\def\th{\theta}
\def\Th{\Theta}
\def\r{\rho}
\def\s{\sigma}
\def\S{\Sigma}
\def\ta{\tau}
\def\o{\omega}
\def\O{\Omega}
\def\pr{\prime}


\def\half{\frac{1}{2}}

\def\goto{\rightarrow}

\def\na{\nabla}
\def\grad{\nabla}
\def\curl{\nabla\times}
\def\div{\nabla\cdot}
\def\pa{\partial}

\def\ll{\left\langle}
\def\rr{\right\rangle}
\def\lb{\left[}
\def\lc{\left\{}
\def\ls{\left(}
\def\lp{\left.}
\def\rp{\right.}
\def\rb{\right]}
\def\rc{\right\}}
\def\rs{\right)}

\def\vac#1{\mid #1 \rangle}


\def\td#1{\tilde{#1}}
\def\check{ \maltese {\bf Check!}}


\def\Tr{{\rm Tr}\,}
\def\det{{\rm det}}


\def\bc#1{\nnindent {\bf $\bullet$ #1} \\ }
\def\ch {$<Check!>$ }
\def\ss {\vspace{1.5cm}}

\begin{titlepage}

\hfill\parbox{5cm} { }

\vspace{25mm}

\begin{center}
{\Large \bf Unbounded Multi-Magnon and Spike}

\vskip 1. cm
 {Bum-Hoon Lee$^a$\footnote{e-mail : bhl@sogang.ac.kr} and
 Chanyong Park$^b$\footnote{e-mail : cyong21@nims.re.kr}}

\vskip 0.5cm

{\it $^a \, $ Department of Physics and Center for Quantum Spacetime, Sogang University, \\
~~Seoul 121-742, Korea \\
 $^b\,$  National Institute for Mathematical Sciences, 385-16 Doryong-dong,
Yuseong-gu, Daejeon 305-340, Korea \\}

\end{center}

\thispagestyle{empty}

\vskip2cm


\centerline{\bf ABSTRACT} \vskip 4mm

\vspace{1cm}

We generalize the one magnon solution in $R \times S^2$ to unbounded $M$ magnon and find
the corresponding solitonic string configuration in the string sigma model.  This
configuration gives rise to
the expected dispersion relation obtained from  the spin chain model in the large 't Hooft
coupling limit. After considering
$(M,M)$ multi-magnon or spike on $R \times S^2 \times S^2$ as a subspace of
$AdS_5  \times S^5$ or $AdS_4 \times CP^3$,
we investigate the dispersion relation and the finite size effect for $(M,M)$ multi-magnon
or spike.

\vspace{2cm}


\end{titlepage}

\renewcommand{\thefootnote}{\arabic{footnote}}
\setcounter{footnote}{0}

\tableofcontents

\section{Introduction}

The AdS/CFT duality \ct{mal1} relates type IIB string theory on
AdS$_5\times$ S$^5$ with ${\cal N} = 4$ superconformal Yang-Mills
(SYM) theory, and it has been celebrated in the last decade as one
of the exact duality between string and gauge theory. Recently
there has been a lot of works devoted towards the understanding of
the worldvolume dynamics of multiple M2-branes \cite{Bagger:2006sk,Aharony:2008ug},
where a new class of conformal invariant 2+1 dimensional field
theories has been found out.


In the development of AdS$_5$/CFT$_4$ duality, an interesting
observation is that the ${\cal N}=4$ SYM theory in planar limit
can be described by an integrable spin chain model where the
anomalous dimension of the gauge invariant operators were found
\ct{Beisert:2005tm,zm,intg1,Beisert:2003xu,intg2,intg3}. It was further noticed
that the string theory is also integrable in the semiclassical
limit \ct{intg4,tsey3,Okamura:2008jm,Hayashi:2007bq, Chen:2006gea}
and the anomalous dimension of the ${\cal N}=4$ SYM can be derived
from the relation among conserved charges of the rotating string
AdS$_5 \times$ S$^5$. In this connection, Hofman and Maldacena
(HM) \ct{Hofman:2006xt} considered a special limit where the
problem of determining the spectrum in both sides becomes rather
simple. The spectrum consists of an elementary excitation known as
magnon that propagates with a conserved momentum $p$ along the
infinitely long \ct{ik0705}-
\ct{Kalousios:2006xy} or the
finitely long
\ct{Ahn:2008sk,Arutyunov:2006gs,Astolfi:2006is,Astolfi:2007uz}
spin chain. In the dual formulation, the most important ingredient
is the semiclassical string solution, which can be mapped to long
trace operator with large energy and large angular momenta.
The integrability of AdS$_5$/CFT$_4$ in the planar limit using the
Bethe ansatz brings us the hope that the recently proposed
AdS$_4$/CFT$_3$ duality will also be solvable by using a similar
ansatz \ct{Minahan:2008hf}. Indeed, in
\cite{Minahan:2008hf,Gaiotto:2007qi,Gaiotto:2008cg,Gromov:2008qe,Arutyunov:2008if,Gomis:2008jt,Stefanski:2008ik}
this has been investigated and many interesting results were
found. The magnon solutions were found in the diagonal SU(2) subsector of ${\bf CP}^3$,
which is similar to
the result obtained from the $AdS_5 \times S^5$ case
\cite{Grignani:2008is}-\ct{Ahn:2008hj}.

In the present paper, we first investigate the dispersion relation of one magnon
on $R \times S^2$
corresponding to the diagonal SU(2) subsector. As will be shown, the dispersion relation
on $R \times S^2$ coming
from $AdS_5 \times S^5$ and $AdS_4 \times CP^3$ has the same form except the different string tension
corresponding to the effective 't Hooft coupling.
By considering the combination of magnons, we generalize one magnon to the configuration which consists
of connected magnons. We will call this configuration as the unbounded multi-magnon or shortly,
multi-magnon. We find the dispersion relation, which
is consistent with the result obtained from the spin chain model in the large 't Hooft coupling limit.
To find the $M$ multi-magnon, we investigate the boundary condition between magnons.
In the infinite size case, the multi-magnon configuration is described by infinitely
differentiable functions with satisfying the equations of motion. For the finite size case,
the multi-magnon configuration is not infinitely differentiable at a joining point of two magnons.
Since the multi-magnon configuration should be a solution of the equation of motion represented
as a second order differential equation, we require that multi-magnon solution becomes differentiable
functions at the second order. This requirement gives rise to a constraint between parameters.
We show that the multi-magnon satisfies the equation of motion at a joining point of two magnons.
Furthermore, the finite size effect are investigated. With the same strategy, multi-spike configuration is also
investigated.

Secondly, the multi-magnon or spike solution on $R \times S^2 \times S^2$,
which is subspace of $AdS_5 \times S^5$
will be considered. In this case, we find out the more general multi-magnon or spike solution, which
describes the combination of $M$ magnon or spike in the first $S^2$ and
$M$ magnon or spike in the second $S^2$
so that we call it $(M,M)$ magnon or spike solution. In Ref.  \ct{Lukowski:2008eq},
using the algebraic curve, it was found
that there exist $(M,N)$ multi-magnon solution. Here, using the string sigma model
we find the string configuration
corresponding to $(M,M)$ multi-magnon. It is still remaining problem to find
the string configuration corresponding to $(M,N)$ multi-magnon.
Here, we will leave that problem as the future works, and calculate the dispersion
relation for $(M,M)$ multi-magnon and spike in the
infinite size case. In addition, the finite size correction following Ref. \cite{Lee:2008ui} is
also calculated.

The rest of the paper is organized as follows. In section 2. we investigate the universal form
of the dispersion relation for a multi-magnon solution on $R \times S^2$,
which is a subspace of $AdS_5 \times S^5$ or $AdS_4\times {\bf CP}^3$. Then, we obtain
the dispersion relation for
multi-magnon consistent with the spin chain's result in the
large 't Hooft coupling limit.
In section 3, we generalize one magnon to the multi-magnon on $R \times S^2$. To obtain
the multi string configuration satisfying the equation of motion, after considering the boundary condition
we find a constraint relation between parameters and the dispersion relation for the multi-magnon
consistent with the spin chain's result in the large 't Hooft coupling limit.
In addition, the finite size effect of the $M$ multi-magnon is also investigated.
In section 4, with the same method we find the dispersion relation and
the finite size effect for $M$ multi-spike.
In section 5, we generalize $M$ multi-magnon and spike on $R \times S^2$ to
$(M,M)$ multi-magnon and spike on $R\times S^2 \times S^2$ and then find
the dispersion relation and the finite size effect.
In section 6, we finish our work with discussion. \\

\section{ One magnon on $R \times S^2$}

According to the AdS/CFT correspondence, the string or the SUGRA
(supergravity) theory on $AdS_m \times {\cal M}$, where ${\cal M}$
is a compact manifold, has a dual SYM theory description on the $AdS$
boundary. If there exists an $S^n$ subspace in the compact manifold ${\cal M}$,
then the string sigma model describing the solitonic open string moving on
$R_t \times S^n$ where
$R_t$ and $S^n$ indicate time in the $AdS$ and an $n$-dimensional sphere
respectively, is reduced to
\be \la{sss}
S = \frac{T}{2} \int d^2 \s \sqrt{- \det h} h^{\a\b} \pa_{\a} X^{\m}
\pa_{\b} X^{\n} G_{\m\n} ,
\ee
where $h^{\a\b}$ and $G_{\m\n}$ are metrics for
the string world sheet and the target space $R_t \times S^n$, respectively.
It is well known that the above string sigma model is classically
integrable. This can be understood by the fact that
through the Polmeyer reduction, the string sigma model is reduced to the
generalized Sine-Gordon theory, which is integrable.
Interestingly, the form of the string action does not depend on the details of
the compact manifold ${\cal M}$ except that $T$ is related to the radius $R_s$ of
$S^n$ as
\be \la{tendef}
T = \frac{R_s^2}{2 \pi} .
\ee
In this section, we will shortly review the universal form of the magnon's dispersion relation
in the $R_t \times S^2$ having the metric
\be
ds^2 = - dt'^2 + R_s^2 \ls d\th^2 + \sin^2 \th d \ph^2 \rs .
\ee
After redefining $t' = R_s t$, the Polyakov string action, \eq{sss}, in the conformal gauge becomes
\be
S = \frac{T}{2} \int d^2 \s \lb (\pa_{\ta} t)^2 - (\pa_{\s} t)^2 - (\pa_{\ta}
\th)^2 + (\pa_{\s} \th)^2 - \sin^2 \th \lc (\pa_{\ta} \ph)^2 -
(\pa_{\s} \ph)^2 \rc \rb ,
\ee
where the string tension is given in \eq{tendef}.

We first consider a point
particle-like solution corresponding to the vacuum solution of the spin chain model
with the
following ansatz
\bea     \la{ppan1}
t &=& \k \tau , \nn
\th &=& \frac{\pi}{2} , \nn
\ph &=& \n \tau ,
\eea
which satisfies all equations of motion for $t$, $\th$ and $\ph$. In the string
sigma model, since the world sheet metric $h_{\a\b}$ is non-dynamical we should
solve the Virasoro constraints. Under the ansatz in \eq{ppan1}, $T_{\ta\s}$
and $T_{\s\s}$ vanish automatically. The remaining one is
$ 0 = T_{\ta \ta} = \k^2 - \n^2 $.
If we consider only positive $\k$ and $\n$, we finally obtain
\be \la{parel}
\k = \n .
\ee
Due to the translation and rotation symmetry in $t$ and $\ph$ coordinate,
there exist two conserved charges
\bea
E &=& T \k, \nn
J &=& T \n .
\eea
Using \eq{parel}, the dispersion relation becomes
\be
E - J = 0,
\ee
which is interpreted as a BPS ground state of the SU(2) spin chain.
According to the AdS/CFT correspondence, the closed string spectrum in the large
N limit is dual to a single trace operator. When considering the open string
which is the half of the closed string with relaxing the boundary condition,
this open string is dual to an operator in the long open spin chain where $J$ corresponds
to the length of the open
spin chain. Furthermore, the isometry
of $S^2$ corresponds to the SU(2) R-symmetry of the boundary gauge theory. The above dispersion relation
implies that the anomalous dimension vanishes, which means that the point like open string
becomes BPS configuration and describes a ground state
of the open spin chain with no impurity or magnon.

To consider the magnon in the open spin chain, we should consider the solitonic
open string in the world sheet\footnote{Here, the solitonic open
string in the world sheet is often called the
giant magnon in the target space and the magnon in the dual open spin chain model.}.
For this, we consider
the following ansatz
\bea     \la{ppan}
t &=& \k \tau , \nn
\th &=& \th(y) , \nn
\ph &=& \n \tau + g(y) ,
\eea
where $y = a \ta + b \s$.
Due to the rotational symmetry in the $\ph$ direction, $g'$ is reduced to
\be
g' = \frac{1}{b^2 - a^2} \ls a \n - \frac{c}{\sin^2 \th} \rs ,
\ee
where the prime means the derivative with respect to $y$ and $c$ is an
integration constant. Here, we choose $b^2  - a^2 > 0$, which describes a magnon.
In other case $a^2 - b^2 >0$,
as will be shown, the solitonic string solution becomes spike.
The equation of motion for $\th$ becomes
\be \la{eqth}
\th'' = - \frac{b^2 \n^2}{(b^2-a^2)^2} \frac{\cos \th}{\sin^3 \th} \ls \sin^4 \th
- \frac{c^2}{b^2 \n^2} \rs .
\ee
Interestingly, this equation can be rederived from the Virasoro constraints represented by
the first order differential equation for $\th$. So we will solve the
Virasoro constraints instead of the second order differential
equation in \eq{eqth}. Here, due to the symmetric property
of $h_{\a\b}$, only three components of $T_{\a\b}$ are independent. Moreover,
the conformal symmetry of the string action, which makes the
energy-momentum tensor traceless, reduces the number of the independent components
to two. For later convenience, we consider the linear combinations of them
\bea
0 &=& T_{\ta\ta} + T_{\s\s} - 2 T_{\ta\s}, \nn
0 &=& T_{\ta\ta} + T_{\s\s} - \frac{a^2 + b^2}{a b}T_{\ta\s}.
\eea
The second Virasoro constraint becomes a relation between parameters
\be \la{2vir}
\k^2 = \frac{c \n}{a} ,
\ee
and the first gives a differential equation for $\th$
\be \la{vth}
\th' = \pm \frac{b \n}{(b^2-a^2) \sin \th} \sqrt{ \ls \sin^2 \th_{max} - \sin^2 \th \rs
\ls \sin^2 \th - \sin^2 \th_{min} \rs} ,
\ee
where
\bea    \la{extr}
\sin^2 \th_{max} &=& \frac{c}{a \n} , \nn
\sin^2 \th_{min} &=& \frac{a c}{b^2 \n} .
\eea
In \eq{vth}, at fixed $\ta$, the plus (or minus) sign implies that the range of $\s$ becomes
$\s_L \le \s \le \bar{\s}$ (or $\bar{\s} \le \s \le \s_R$), where $\th_{min} = \th(\bar{\s})$
and $\s_L$ or $\s_R$ means the left or right end satisfying
$\th_{L,max} = \th_{max} = \th (\s_L)$ or $\th_{R,max} = \th_{max} = \th (\s_R)$, respectively.
Note that the differentiation of \eq{vth} with respect to $y$ gives rise to
the equation of motion for $\th$ in \eq{eqth}.
To consider the giant magnon having infinite angular momentum, which is called
the infinite size limit and corresponds
to the magnon in the infinitely long open spin chain, we should set
$\sin \th_{max} = 1$. Then, from \eq{2vir} and \eq{extr}, some relations between
parameters are given by
\bea    \la{rvar}
a &=& \frac{c}{\n} , \nn
\k &=&  \n .
\eea

In this infinite size limit, \eq{vth} is rewritten as
\be \la{theq1}
\th' = \pm \frac{\n \cos \th \sqrt{\sin^2 \th - \sin^2 \th_{min} }}
{b \cos^2 \th_{min} \sin \th} ,
\ee
with
\be
\sin^2 \th_{min} = \frac{c^2}{b^2 \n^2} .
\ee
Note that using \eq{rvar} $\sin \th_{min}$ can be rewritten as $\frac{a}{b}$,
which implies that the minimum value of $\th$ is
determined by the ratio between $a$ and $b$ only.
Using the above results, the conserved charges are given by
\bea
E &=& 2 T  \int_{\th_{min}}^{\pi/2} d \th
\frac{\cos^2 \th_{min} \sin \th}{\cos \th \sqrt{\sin^2 \th - \sin^2 \th_{min}}} , \nn
J &=& 2 T \int_{\th_{min}}^{\pi/2} d\th \frac{\sin \th (\sin^2 \th
- \sin^2 \th_{min})}{\cos \th \sqrt{\sin^2 \th - \sin^2 \th_{min}}} .
\eea
Another important quantity is the world sheet
momentum $p$, which corresponds to the angle difference $\D \ph =p$ in the target
space,
\bea \la{angs}
\D \ph &\equiv& \int d \ph =   2
\int_{\th_{min}}^{\pi/2} d \th
      \frac{\cos \th \sin \th_{min}}{\sin \th \sqrt{ \sin^2 \th - \sin^2 \th_{min} }} .
\eea
Using these, the dispersion relation for one giant magnon corresponding
to the solitonic string moving on $R_t \times S^2$ can be universally described by
\be \la{diss2}
E - J = 2 T \left| \sin \frac{p}{2} \right|
\ee
in the large 't Hooft coupling limit, where the string tension $T$ is also large.
For examples, in the $AdS_5 \times S^5$ case \ct{ik0705},
since $R_s^2 = \sqrt{\l}$, the dispersion relation reduced to
\be
E - J = \frac{\sqrt{\l}}{\pi} \left| \sin \frac{p}{2} \right| .
\ee
For the the $AdS_4 \times {\bf CP}^3$ case \cite{Lee:2008ui},
the radius of $S^2$ is $R_s^2 =  \pi \sqrt{2 \l}$, so the dispersion relation
becomes
\be
E - J = \sqrt{2 \l} \left| \sin \frac{p}{2} \right| .
\ee

In the gauge theory side, the dual description for a giant magnon comes from
the integrable spin chain model, in which the dispersion relation for one magnon
is given by
\be
E - J = \sqrt{ 1 +  4 T^2  \sin^2 \frac{p}{2}  } .
\ee
In the large 't Hooft coupling limit, this dispersion relation becomes
\eq{diss2}.

\section{Multi-magnon on $R \times S^2$}

From now on, we consider unbounded $M$ multi-magnon,
where $M$ is the number of the magnon.
For this, we parameterize the $\s$ range of the $i$-th magnon as
\be
 -L + L \sum_{i=1}^{m-1} l_i  \le \s_m \le -L + L \sum_{i=1}^{m} l_i ,
\ee
where $\sum_{i=1}^{M} l_i = 2$ so that the total distance of $M$ magnon
in the $\s$ direction is $2L$. The range of $\ta$ is parameterized as $- L < \ta < L$.
Then, the $M$ multi-magnon can be described by the same action in the previous section
\be
S = \frac{T}{2} \int_{-L}^{L} d \ta \int_{-L}^{L} d \s
\sqrt{- \det h} h^{\a\b} \pa_{\a} X^{\m}
\pa_{\b} X^{\n} G_{\m\n} ,
\ee
with the following ansatz
\bea     \la{mpar}
t &=& \k \tau , \nn
\th &=& \sum_{m=1}^{M} \th_{m}(y_m) , \nn
\ph &=& \sum_{m=1}^{M} \ph_m = \sum_{m=1}^{M} \n_{m} \tau + g_{m}(y_m),
\eea
where the subscript $m$ implies the $m$-th magnon and $y_m = a_m \ta + b_m \s_{m}$.

Due to the rotational symmetry in the $\ph$ direction,
the equation of motion for $\ph_m$ reduces to
\be \la{eqph}
g_m' = \frac{1}{b_m^2 - a_m^2} \ls  a_m \n_m  - \frac{c_m}{\sin^2 \th_m} \rs ,
\ee
where $c_m$ is an integration constant. The energy-momentum tensor for $m$-th magnon becomes
\bea
T^m_{\ta \ta} + T^m_{\s \s}
&=& \frac{T}{2}  \lb - \k^2 + (a_m^2 + b_m^2) (\th_m')^2
 + (\n_m + a_m g_m')^2 + b_m^2 (g_m')^2 \rb , \nn
T^m_{\ta \s}
&=& \frac{T}{2} a_m b_m (\th_m')^2 + b_m (\n_m + a_m g_m') g_m' .
\eea
From these, the Virasoro constraints can be rewritten as the more simple form like
\bea
0 &=&   T^m_{\ta \ta} + T^m_{\s \s} - \frac{a_m^2 +  b_m^2}{a_m b_m} T^m_{\ta \s} , \nn
0 &=&  T^m_{\ta \ta} + T^m_{\s \s} - 2 T^m_{\ta \s}  .
\eea
Since the energy-momentum tensor vanishes at all $\ta$ and $\s$, the above Virasoro constraints
should be applied to all $M$ magnon.
The first constraint gives rise to
\be \la{timev}
\k^2 = \frac{\n_m  c_m}{a_m} ,
\ee
and the second constraint
\be \la{virth}
\th_m' = \pm \frac{b_m \n_m}{(b_m^2-a_m^2) \sin \th_m}
\sqrt{( \sin^2 \th_{m,max} - \sin^2 \th_m)
( \sin^2 \th_m - \sin^2 \th_{m,min})} ,
\ee
where
\bea    \la{mm}
\sin^2 \th_{m,max} &=& \frac{c_m}{a_m \n_m} , \nn
\sin^2 \th_{m,min} &=& \frac{a_m c_m}{b_{m}^2 \n_m} .
\eea

For the multi-magnon configuration to be a solution,
$m$-th magnon solution should be smoothly connected to $(m-1)$-th
and $(m+1)$-th magnon. This implies that values of $\th$ and $\ph$ for the right end of $m$-th
and the left end of $(m+1)$-th magnon are smoothly connected. First, to satisfy this
boundary condition, $\th^R_{m,max}=\th_{m,max}$ should be same as $\th^L_{m+1,max}=\th_{m+1,max}$,
which implies that $\th_{m}$ for all $m$-th magnon has the same maximum value
\be
\th_{m,max} = \th_{max} \ {\rm for \  all} \ m .
\ee
Using this, $c_m$ can be rewritten in terms of other parameters
\be \la{con22}
c_m = a_m \n_m \sin^2 \th_{max} .
\ee
With \eq{timev}, this relation means that $\n_m$ is constant like
$\n_m = \n$ for all $m$.
Because of the smoothness of $\th$ at $\th_{max}$, $\s_m$
is represented
as an integration with respect to $\th$
\be
\s_m - \s^L_m = \int_{\th^L_{m,max}}^{\th_m} \frac{d\th_m}{b_m \th_m'} ,
\ee
where $\s^L_m$ and $\th^L_{m,max}$ are the values of $\s_m$ and $\th_m$ at the left end of the
$m$-th magnon.

In \eq{virth}, $\th'$ always vanishes at $\th_{max}$ so that $\frac{\pa \th}{\pa \s}$ is a smooth
function of $\s$ at fixed $\ta$.
To investigate the smoothness for the higher derivatives of $\th$,
we write several terms by using the chain rule
\bea    \la{hde}
\th^{(2)} &=& \frac{b_m^2}{2} \frac{\pa}{\pa \th} \th'^2 , \nn
\th^{(3)} &=& \frac{b_m^3}{2} \th' \ \frac{\pa^2}{\pa \th^2} \th'^2 , \nn
\th^{(4)} &=& \frac{b_m^4}{4} \frac{\pa}{\pa \th} \th'^2 \ \frac{\pa^2}{\pa \th^2} \th'^2, \nn
\th^{(5)} &=& b_m^5 \th' \ \ls \frac{1}{4}  \frac{\pa^2}{\pa \th^2} \th'^2 \ \frac{\pa^2}{\pa \th^2} \th'^2
+ \frac{3}{4}  \frac{\pa}{\pa \th} \th'^2 \ \frac{\pa^3}{\pa \th^3} \th'^2
+ \half \th'^2 \ \frac{\pa^4}{\pa \th^4} \th'^2 \rs , \nn
\cdots \ ,
\eea
where $\th^{(n)}$ means $\frac{\pa^n}{\pa \s^n} \th$.
Since $\th'$ vanishes at $\th_{max}$, $\th^{(n)}$ for odd $n$ becomes automatically zero.
For even $n$ case, at $\th^R_{m,max}$ where $m$-th and $(m+1)$-th magnon are joined
$\th^{(n)}$'s values of $m$-th and $(m+1)$-th magnon are different, so in general $\th$ is not differentiable.
Since multi-magnon configuration should be a solution of the equation of motion represented as
a second order differential equation, we have to require that at least $\th^{(n)}$ with $n \le 2$ is
differentiable. As previously mentioned, when $\th^R_{m,max}= \th^L_{m+1,max} = \th_{max}$,
$\th^{(0)}$ is smooth. As will be shown, the smoothness of $\th^{(2)}$ determines the minimum value of $\th$
in terms of other parameters, $a_m$, $b$ and $\th_{max}$.
This is a story of the multi-magnon with a finite size. In the infinite size case with
$\th_{max} = \pi/2$, since $\th'$ and $\frac{\pa^2 }{\pa \th^2} \th'^2$ are proportional to $\cos \th$,
$\th^{(n)}$ for all $n$ vanishes at $\th_{max}$. So $\th$ in the infinite size case becomes an infinitely
differentiable function.

Before investigating the multi-magnon with finite size, we consider the infinite size case.
In this case, \eq{virth} and \eq{eqph} are reduced to simpler forms
\bea
\th_m' &=& \pm \frac{b_m \n \cos \th_m}{(b_m^2-a_m^2) \sin \th_m}
\sqrt{ \sin^2 \th_m - \sin^2 \th_{m,min}} , \nn
g_m' &=& \frac{a_m \n}{b_m^2 - a_m^2}
 \ls 1  - \frac{1}{\sin^2 \th_m} \rs
\eea
with
\bea
\k &=& \n , \nn
\sin^2 \th_{max} &=& 1 , \nn
\sin^2 \th_{m,min} &=& \frac{a_m^2}{b_m^2 } .
\eea
Using these results, the conserved charges for the multi-magnon are given by
\bea
E &=& 2 T  \sum_{m=1}^{M} \int_{\th_{m,min}}^{\pi/2} d \th_m
\frac{\cos^2 \th_{m,min} \sin \th_m }{\cos \th_m \sqrt{\sin^2 \th_m
 - \sin^2 \th_{m,min}}} ,\nn
J &=& 2 T \sum_{m=1}^{M} \int_{\th_{{m},min}}^{\pi/2} d\th_{m}
\frac{\sin \th_{m}
(\sin^2 \th_{m} - \sin^2 \th_{{m},min})}{\cos \th_{m} \sqrt{\sin^2 \th_{m}
- \sin^2 \th_{{m},min}}} .
\eea
The angle difference corresponding to the world momentum of the $m$-th magnon is given by
\bea \la{mang}
p_{m}
= 2 \int_{\th_{{m},min}}^{\pi/2} d\th_{m} \frac{\cos \th_{m} \sin \th_{{m},min}}
{\sin \th_{m} \sqrt{\sin^2 \th_{m} - \sin^2 \th_{{m},min}}}  .
\eea
Calculating these integrations, we obtain
\bea \la{madwm}
\sin \th_{{m},min} &=& \cos \ls \frac{p_{m}}{2} \rs  ,
\eea
which can be rewritten as $p_m = \pi - 2 \th_{{m},min}$.
This result implies that the considered multi-magnon configuration allows for each magnon to
have different world sheet momentum $p_m$. Therefore, the dispersion relation for $M$ multi-magnon becomes
\bea    \la{gedr}
E - J &=& 2 T \sum_{m=1}^{M}
\left| \sin \frac{p_{m}}{2} \right| .
\eea
Comparing this result with one obtained from the spin chain model
\be \la{fmm}
E - J = \sqrt{1 + 4 T^2 \sum_{m=1}^{M} \left| \sin \frac{p_{m}}{2} \right|^2 },
\ee
in the large 't Hooft coupling limit \eq{fmm} becomes \eq{gedr}.

For the multi-magnon with the finite size,
the equations for this string configuration are
\bea
\th_m' &=& \pm \frac{b_m \n}{(b_m^2-a_m^2) \sin \th_m}
\sqrt{( \sin^2 \th_{max} - \sin^2 \th_m)
( \sin^2 \th_m - \sin^2 \th_{m,min})} , \nn
g_m' &=& \frac{a_m \n}{b_m^2 - a_m^2}
 \ls 1 - \frac{ \sin^2 \th_{max}}{\sin^2 \th_m} \rs ,
\eea
where
\bea
\sin^2 \th_{m,min} &=& \frac{a_m^2}{b_{m}^2} \sin^2 \th_{max} .
\eea
\eq{timev} and \eq{con22} give the following relation
\be
 \k^2 = \n^2 \sin^2 \th_{max} .
\ee
In this case, $\th_m'$, $g_m'$ and $g_m''$ are automatically zero at $\th_{m,max}=\th_{max}$
where $\th$ is smooth.
$\th_m''$ at $\th_{max}$ becomes
\be
\lp \th_m'' \right|_{\th = \th_{max}} = - \frac{\n^2}{b_m^2 - a_m^2} \sin \th_{max}
\cos \th_{max} ,
\ee
which satisfy the equations of motion in \eq{eqth} with $a_m$ and $b_m$ instead of $a$ and $b$.
Therefore, the differentiability of $\th^{(2)}$ requires that $b_m^2 - a_m^2$ be a constant
independent of $m$. This proves the existence of the $M$ multi-magnon solution.
Following Ref. \ct{Lee:2008ui}, the dispersion relation with the finite size effect for the $M$ multi-magnon is
\be \la{fs}
E - J = 2 T \sum_{m=1}^{M} \left(\left|  \sin
\frac{p_m}{2} \right| - 4 \left| \sin^3 \frac{p_m}{2} \right| e^{- E_m /
( T \left| \sin \frac{p_m}{2} \right| )}\right) ,
\ee
where $E_m$ is the energy of the $m$-th magnon.

\section{Multi-spike on $R \times S^2$}

Like the multi-magnon, we can also consider a multi-spike solution, for which we set $a_m^2 - b_m^2$
positive
and interchange the roles of $\sin_{m,max}$ and $\sin_{m,min}$ in \eq{virth} and \eq{mm}
\bea \la{virth2}
\th_m' &=& \pm \frac{b_m \n_m}{(a_m^2-b_m^2) \sin \th_m}
\sqrt{( \sin^2 \th_{m,max} - \sin^2 \th_m)
( \sin^2 \th_m - \sin^2 \th_{m,min})} , \nn
g_m' &=& \frac{1}{a_m^2 - b_m^2} \ls \frac{c_m}{\sin^2 \th_m}  -   a_m \n_m  \rs ,
\eea
where
\bea    \la{mm2}
\sin^2 \th_{m,max} &=& \frac{a_m c_m}{b_{m}^2 \n_m} , \nn
\sin^2 \th_{m,min} &=&  \frac{c_m}{a_m \n_m} .
\eea
and the constraint equation becomes
\be
\k^2 = \frac{\n_m  c_m}{a_m}  .
\ee
For the smooth solution, we set $\sin \th_{m,max} = \sin \th_{max}$ for all $m$. Then, $\n_m$ can be
rewritten as
\be
 \n_m = \frac{a_m c_m}{b_{m}^2 \sin^2 \th_{max}},
\ee
where $c_m = \k \sin \th_{max} b_m$.
Inserting this to the last relation in \eq{mm2}, $\sin^2 \th_{m,min}$ becomes
\be
\sin^2 \th_{m,min} = \frac{b_m^2}{a_m^2} \sin^2 \th_{max} .
\ee
Like the multi-magnon case, in the infinite size limit $\th_m$ and $\ph_m$
are infinitely differentialbe
functions and there is no restriction to the value of $\sin \th_{m,min}$. For this case,
the conserved charges
and the angle difference for $M$ multi-spike are given by
\bea
E &=& 2 T \sum_{m=1}^{M} \int_{\th_{m,min}}^{\pi/2} d \th_m
\frac{\cos^2 \th_{m,min} \sin \th_{m}}{\sin \th_{m,min} \cos \th_m }
\frac{1}{\sqrt{\sin^2 \th_m - \sin^2 \th_{m,min}}} , \nn
J &=& 2 T  \sum_{m=1}^{M} \int_{\th_{m,min}}^{\pi/2} d\th_m
\frac{\sin \th_m  \cos \th_m}{\sqrt{\sin^2 \th_m
- \sin^2 \th_{m,min}}} , \nn
\D \ph
&=& 2 \sum_{m=1}^{M} \int_{\th_{m,min}}^{\pi/2} d \th_m
\frac{\sin^2 \th_m - \sin^2 \th_{m,min} }{\sin \th_{m,min} \cos \th_m \sin \th_m}
\frac{1}{\sqrt{\sin^2 \th_m - \sin^2 \th_{m,min}}},
\eea
From these, the dispersion relation for the $M$ multi-spike is
\bea
E - T \D \ph &=& 2 T \sum_{m=1}^{M} \td{\th}_m , \nn
J_m &=& 2 T \sin \td{\th}_m  ,
\eea
where $\D \ph = \sum_{m=1}^{M} \D \ph_m$ and $\td{\th}_m = \pi/2 - \th_{m,min}$.

In the finite size case,
the differentiability of $\th^{(n)}$ for $n \le 2$ gives a relation for $\th_{m,min}$
\be
\frac{\sin \th_{m,min}}{\sin \th_{m+1,min}} = \frac{ a_m^2 - b_m^2 }{  a_{m+1}^2 - b_{m+1}^2  } ,
\ee
which guarantees the existence of the solution. In the infinite size case, since there is no this kind
of constraint $a_m$ and $b_{m}$ can have arbitrary values.
The dispersion relation with the finite size effect for this $M$ multi-spike
can be easily calculated by following Ref. \ct{Lee:2008ui}
\bea
E- T \D \ph &\approx&  \sum_{m=1}^{M} \lb 2 T \td{\th}_m
- \lc \ls 4 - \frac{4}{\sqrt{1- (J_m/2T)^2}} - \frac{J_m^2}{2T^2} \rs J_m \rp \rp \nn
&& \lp  \lp  + \ls 8 - \frac{8}{\sqrt{1- (J_m/2T)^2}}
+ \frac{J_m^2}{T^2} \rs E_m \rc \ e^{-2E_m/J_m}  \rb ,
\eea
where $E_m$ and $J_m$ are the energy and angular momentum for $m$-th spike.

\section{Multi-magnon and spike on $R \times S^2 \times S^2$}

In this section, we will generalize the multi-magnon and spike configuration on $R \times S^2$ to
ones on $R \times S^2 \times S^2 $, which corresponds to the $SU(2) \times SU(2)$ subsector of
$SU(4)$ R-symmetry in the boundary gauge theory. Here, the background space, $R \times S^2 \times S^2$,
can be obtained as a subspace of $AdS_5 \times S^5$ or
$AdS_4 \times CP^3$. In the later case, the calculation and the result
are similar to those on $AdS_5 \times S^5$ with a constraints $\ps=$constant, $\xi=\pi/4$
(see details Ref. \ct{Lee:2008ui}).

To obtain $R \times S^2 \times S^2$, we first consider $R \times S^5$, which is
the subspace of $AdS_5 \times S^5$ at the center of $AdS_5$. $S^5$ can be
represented by a hypersurface in the six-dimensional flat space by imposing a constraint
\be
R^2 = \sum_{i=1}^{6} X_i^2 .
\ee
The parameterization satisfying the above constraint is given by
\bea    \la{coordan}
X_1 &=& R \sin \xi \sin \th_1 \sin \ph_1  ,\nn
X_2 &=& R \sin \xi \sin \th_1 \cos \ph_1   ,\nn
X_3 &=& R \sin \xi \cos \th_1    ,\nn
X_4 &=& R \cos \xi \sin \th_2 \sin \ph_2   ,\nn
X_5 &=& R \cos \xi \sin \th_2 \sin \ph_2   ,\nn
X_6 &=& R \cos \xi \cos \th_2  .
\eea
Then, the flat metric in the six-dimensional space reduces to one for $S^5$
\be \la{mets5}
ds^2 = R^2 \lb - dt^2 + d\xi^2 + \sin^2 \xi ( d\th_1^2 + \sin^2 \th_1 d\ph_1^2)
 + \cos^2 \xi ( d\th_2^2 + \sin^2 \th_2 d\ph_2^2) \rb ,
\ee
where we insert a time direction.
The string action moving in this target space becomes
\begin{eqnarray}
S &=& \frac{T}{2} \int d^2 \sigma  \left[
- \partial_{\alpha} t \partial^{\alpha} t + \partial_{\alpha} \xi \partial^{\alpha} \xi
+ \sin^2 \xi (\partial_{\alpha} \theta_1 \partial^{\alpha} \theta_1
+ \sin^2 \theta_1 \partial_{\alpha} \phi_1 \partial^{\alpha} \phi_1) \rp \nonumber \\
&& \left. \qquad \qquad  + \cos^2 \xi (\partial_{\alpha} \theta_2 \partial^{\alpha}
\theta_2
+ \sin^2 \theta_2 \partial_{\alpha} \phi_2 \partial^{\alpha} \phi_2) \right] .
\end{eqnarray}
To obtain the string solution moving in $R \times S^2 \times S^2$,
we investigate the equation of motion for $\xi$
\bea    \la{eqxi}
0 &=& \partial^{\alpha} \partial_{\alpha} \xi
- \sin \xi \cos \xi \lb \partial_{\alpha} \theta_1 \partial^{\alpha} \theta_1
+ \sin^2 \theta_1 \partial_{\alpha} \phi_1 \partial^{\alpha} \phi_1 \rp \nn
&&  \qquad  \qquad  \qquad  \qquad \lp - \partial_{\alpha} \theta_2 \partial^{\alpha}
\theta_2
- \sin^2 \theta_2 \partial_{\alpha} \phi_2 \partial^{\alpha} \phi_2 \rb .
\eea
For $\xi=0$ or $\pi/2$, the above equation is automatically satisfied and the
metric in \eq{mets5} reduces to $R \times S^2$, which has been already investigated
in the previous section. To satisfy \eq{eqxi}, we will consider the case $\xi=$constant
but $\xi=0$ or $\pi/2$, in which
$\th_1 = \th_2$ and $\ph_1 = \ph_2$ satisfy \eq{eqxi}. So the solitonic string with these
constraints describes
a $(M,M)$ multi-magnon, which implies $M$ multi-magnon in each $S^2$.
Note that in the $AdS_4 \times {\bf CP}^3$ case, the parameterization in \eq{coordan}
satisfy the ${\bf CP}^3$ constraint
only at $\xi=\pi/4$ \cite{Lee:2008ui}.

Since all calculation is very similar to the previous one, we skip the details. To consider
the $(M,M)$ multi-magnon and
spike on $R \times S^2 \times S^2$, the appropriate
ansatz is given by
\bea    \la{ans2s2}
t &=& \sum_{m=1}^{M}  \ls \sin^2 \xi  +  \cos^2 \xi  \rs^{1/2} \k \ta , \nn
\th_{1,m} &=& \th_{2,m} = \th_m  \ls y_{m} \rs , \nn
\ph_{1,m} &=& \ph_{2,m}  = \n_{m} \ta + f_{m} \ls y_{m} \rs   ,
\eea
where the subscript $1$ and $2$ in the last two equations represent the first and second $S^2$
sphere and and
$y_{m} = a_{m} \ta + b_{m} \s_{m}$. Note that in the first line in \eq{ans2s2} we write
$\sin^2 \xi  +  \cos^2 \xi$
explicitly for later convenience.
Then, in the infinite size limit the dispersion relation for each sphere is given by
\bea
E_1 - J_1 &=& 2T \sin^2 \xi \sum_{m=1}^{M} \left| \sin \frac{p_m}{2} \right| , \nn
E_2 - J_2 &=& 2T \cos^2 \xi \sum_{m=1}^{M} \left| \sin \frac{p_m}{2} \right|
\eea
where
\bea
E_1 &=& 2 T \sin^2 \xi  \sum_{m=1}^{M} \int_{\th_{m,min}}^{\pi/2} d \th_m
\frac{\cos^2 \th_{m,min} \sin \th_{m}}{\cos \th_{m} \sqrt{\sin^2 \th_{m}
 - \sin^2 \th_{m,min}}}  , \nn
E_2 &=&  2 T \cos^2 \xi \sum_{m=1}^{M} \int_{\th_{m,min}}^{\pi/2} d \th_{m}
\frac{\cos^2 \th_{m,min} \sin \th_{m}}{\cos \th_{m} \sqrt{\sin^2 \th_{m}
- \sin^2 \th_{{m},min}}} , \nn
J_1 &=& 2 T \sin^2 \xi \sum_{m=1}^{M} \int_{\th_{{m},min}}^{\pi/2} d\th_{m}
\frac{\sin \th_{m}
(\sin^2 \th_{m} - \sin^2 \th_{{m},min})}{\cos \th_{m} \sqrt{\sin^2 \th_{m}
- \sin^2 \th_{{m},min}}} , \nn
J_2 &=& 2 T \cos^2 \xi \sum_{n=1}^{N} \int_{\th_{{m},min}}^{\pi/2} d\th_{m}
\frac{\sin \th_{m}
(\sin^2 \th_{m} - \sin^2 \th_{{m},min})}{\cos \th_{m} \sqrt{\sin^2 \th_{m}
- \sin^2 \th_{{m},min}}} , \nn
p_{m} &=& 2 \int_{\th_{{m},min}}^{\pi/2} d\th_{m} \frac{\cos \th_{m} \sin \th_{{m},min}}
{\sin \th_{m} \sqrt{\sin^2 \th_{m} - \sin^2 \th_{{m},min}}} .
\eea
Therefore, the total dispersion relation for $(M,M)$ multi-magnon in $AdS_5 \times S^5$ becomes
\bea
E - J &=& 2T \sin^2 \xi \sum_{m=1}^{M} \left| \sin \frac{p_m}{2} \right|
+ 2T \cos^2 \xi \sum_{m=1}^{M} \left| \sin \frac{p_m}{2} \right| \la{uvdis} \\
&=&  \sum_{m=1}^{M} \frac{\sqrt{\l_{1}}}{\pi} \ \left| \sin \frac{p_{m}}{2} \right|
+  \sum_{m=1}^{M} \frac{\sqrt{\l_{2}}}{\pi}  \left| \sin \frac{p_{m}}{2} \right| ,
\eea
where $\l_1= \sqrt{\l \sin^4 \xi}$ and $\l_2 = \sqrt{\l \cos^4 \xi }$. Due to the different radius of
two $S^2$, the effective 't Hooft coupling in each sphere has different value. If we consider the diagonal
subgroup of $SU(2) \times SU(2)$, then it reduces to the previous result obtained in the
$R \times S^2$ case. Note that \eq{uvdis} is a universal form of the dispersion relation for
$(M,M)$ multi-magnon on $R \times S^2 \times S^2$. So to describe the $(M,M)$ multi-magnon
moving in the $AdS_4 \times CP^3$, we should set $\xi = \pi/4$ and $T = \sqrt{2 \l}/2$
so that the dispersion relation becomes
\be
E - J =  \sum_{m=1}^{M}  \sqrt{2 \l}\ \left| \sin \frac{p_{m}}{2} \right|  ,
\ee
which is the same form obtained from the $R \times S^2$ case. Note that the dispersion relation
for the each sphere
is
\be
E_i - J_i =  \sum_{m=1}^{M}  \frac{\sqrt{2 \l}}{2} \ \left| \sin \frac{p_{m}}{2} \right|  ,
\ee
where $i$ means $i$-th sphere.

The finite size effect of $(M,M)$ multi-magnon coming from $AdS_5 \times S^5$ background is given by
\bea
E - J &=& \frac{\sqrt{\l_{1}}}{\pi} \sum_{m=1}^{M} \left(\left|  \sin
\frac{p_m}{2} \right| - 4 \left| \sin^3 \frac{p_m}{2} \right| e^{- 2 \pi E_m /
( \sqrt{\l_{1}} \left| \sin \frac{p_m}{2} \right| )}\right)  \nn
&& \quad + \frac{\sqrt{\l_{2}}}{\pi} \sum_{m=1}^{M} \left(\left|  \sin
\frac{p_m}{2} \right| - 4 \left| \sin^3 \frac{p_m}{2} \right| e^{- 2 \pi E_m /
( \sqrt{\l_{2}} \left| \sin \frac{p_m}{2} \right| )}\right) .
\eea
In the $AdS_4 \times CP^3$ case, the finite size effect for the $i$-th sphere is
\bea \la{fcp3}
E_i - J_i &=& \frac{\sqrt{2 \l}}{2} \sum_{m=1}^{M} \left(\left|  \sin
\frac{p_m}{2} \right| - 4 \left| \sin^3 \frac{p_m}{2} \right| e^{- 2 E_m /
( \sqrt{2 \l} \left| \sin \frac{p_m}{2} \right| )}\right)  .
\eea
So the total finite size effect becomes twice of \eq{fcp3}.

In the infinite size limit, the dispersion relation of the $(M,M)$ multi-spike on the $i$-th sphere becomes
\bea
E_i - T_i \sum_{m=1}^{M} \D \ph_{m}
&=& 2 T_i  \sum_{m=1}^{M} \td{\th}_{m} , \nn
J_i &=& 2 T_i  \sum_{m=1}^{M} \sin \td{\th}_{m}  ,
\eea
where $\td{\th}_{m} = \pi/2 - \th_{m,min}$ and $\lc T_1,  \ T_2 \rc = \lc T \sin^2 \xi, \ T \cos^2 \xi \rc$
with $T = \frac{\sqrt{\l}}{2 \pi}$ for $AdS_5 \times S^5$ case and
$T =\frac{\sqrt{2 \l}}{2}$ and $\xi = \pi/4$ for $AdS_4 \times CP^3$ case. The finite size effect for
$(M,M)$ multi-spike on the $i$-th sphere of $AdS_5 \times S^5$ becomes
\bea
E_i - T_i \D \ph &\approx& \sum_{m=1}^{M} \lb  2 T_i \td{\th}_m
- \lc \ls 4 - \frac{4}{\sqrt{1 - (J_{i,m}/2T_i)^2}} - \frac{J_{i,m}^2}{2T_i^2} \rs J_{i,m} \rp \rp \nn
&& \lp  \lp  + \ls 8 - \frac{8}{\sqrt{1- (J_{i,m}/2T_i)^2}}
+ \frac{J_{i,m}^2}{T_i^2} \rs E_{i,m} \rc \ e^{-2E_{i,m}/J_{i,m}}  \rb  ,
\eea
where
\bea
E_{i,m}  &=&  T_i \ls  \D \ph_{m} + 2 \td{\th}_{m} \rs ,\nn
J_{i,m} &=& 2 T_i  \sin \td{\th}_{m} .
\eea


\section{Discussion}

At first, we found a universal form of the dispersion relation for a magnon solution
on $R \times S^2$. Only the difference between the magnon moving in $AdS_5 \times S^5$
and in $AdS_4 \times {\bf CP}^3$ is the different radius of $S^2$, which is represented
as a different string tension corresponding to the 't Hooft coupling. The universality of
the dispersion relation can be easily extended to $R \times S^n$ case. Here, the universal form
on $R \times S^2 \times S^2$ was also investigated. For example, \eq{uvdis} is the
universal form of the dispersion relation
for the multi-magnon on $R \times S^2 \times S^2$.

Secondly, we investigated the multi-magnon configuration consisting of $M$ magnons.
this corresponds to the unbounded multi-magnon or shortly, multi-magnon
in spin chain model. In the infinite size limit, the dispersion relation of multi
magnon gives the same result obtained from the spin chain model in the large 't Hooft coupling limit.
In the finite size case, though the multi-magnon configuration is not infinitely differentiable
at $\th_{max}$, by requiring the differentiability of $\th^{(2)}$ we found the multi
magnon configuration with the finite size satisfying the equation of motion and calculated
the finite size effect. With the same method, we also investigated the string configuration corresponding
to the multi-spike and found the dispersion relation and the finite size effect.
Furthermore, we generalized the multi-magnon solution on $R \times S^2$ to
$(M,M)$ multi-magnon solution on $R \times S^2 \times S^2$. For the case of the multi-magnon moving
in $AdS_5 \times S^5$, the radius of $S^2$ depends on the value of $\xi$ so that $(M,M)$ multi
magnon describes the combination of two $M$ multi-magnons on each sphere
with the effectively different 't Hooft coupling.

The remaining interesting problem is to find the $(M,N)$ type multi-magnon or spike configuration from
the string sigma model. Due to the constraint in \eq{eqxi}, it is not easy to find $(M,N)$
multi-magnon configuration. We hope to find those string configuration in the next work.

\vspace{1cm}

{\bf Acknowledgement}

We would like to thank to Changrim Ahn, P. Bozhilov,  Corneliu Sochichiu, Dongsu Bak and
Ki-Myeong Lee for valuable discussion. B.-H. Lee was was supported by the Science Research
Center Program of the Korean Science and Engineering Foundation through the Center for Quantum
SpaceTime (CQUeST) of Sogang University with grant number R11-2005-021.
C. Park was supported by the Korea Research Council of Fundamental Science and Technology (KRCF).

\vspace{1cm}


\end{document}